\documentclass[prc,superscriptaddress,unsortedaddress,
twocolumn,showpacs,preprintnumbers,amsmath,amssymb]{revtex4}
\usepackage[dvips]{graphicx}
\usepackage{amsmath}
\usepackage{amssymb}
\usepackage{times}
\usepackage{color}
\def\mbold#1{\mbox{\boldmath $#1$}}
\def\non{\nonumber}

\begin{document}

\title{Systematic description of $^6$Li($n$, $n'$)
$^6$Li$^*$ $\to$ $d$ + $\alpha$
 reactions with the microscopic coupled-channels method}   

\author{T. Matsumoto}
\email[Electronic address: ]{tmatsumoto@nucl.sci.hokudai.ac.jp}
\affiliation{Meme Media Laboratory, Hokkaido University, Sapporo
060-8628, Japan} 

\author{D. Ichinkhorloo}
\affiliation{Division of Physics, Graduate School of Science, Hokkaido
University, Sapporo 060-0810, Japan} 

\author{Y. Hirabayashi}
\affiliation{Information Initiative Center, Hokkaido
University, Sapporo 060-0810, Japan} 

\author{K. Kat\=o}
\affiliation{Division of Physics, Graduate School of Science, Hokkaido
University, Sapporo 060-0810, Japan} 

\author{S. Chiba}
\affiliation{Japan Atomic Energy Agency (JAEA), Tokai, Naka, Ibaraki
319-1195, Japan} 

\date{\today}

\begin{abstract}
 We investigate $^6$Li($n$, $n'$)$^6$Li$^*$ $\to$ $d$ +
 $\alpha$ reactions by using the  continuum-discretized coupled-channels
 method with the complex Jeukenne-Lejeune-Mahaux effective
 nucleon-nucleon interaction. In this study, the $^6$Li nucleus is
 described as a $d$ + $\alpha$ cluster model. The calculated elastic
 cross  sections for incident energies between 7.47 and 24.0 MeV are
 good agreement with experimental data. Furthermore, we show the neutron
 spectra to $^6$Li breakup states measured at selected angular points
 and incident energies can be also reproduced systematically.
\end{abstract}


\maketitle

\section{Introduction}
\label{sec:intro}

The $^6$Li nucleus is known to have a well developed $d$ + $\alpha$
cluster structure and the binding energy is very small $1.47$ MeV from
the $d$-$\alpha$ breakup threshold. Because of those features, a breakup
process of $^6$Li into $d$ and $\alpha$ is one of the significant
mechanisms of the reaction, and the accurate description is required. 
As one of the most reliable methods for treating breakup processes, 
the continuum-discretized coupled channels (CDCC)
method~\cite{Kamimura} has been proposed and successfully
applied to analyses of three-body breakup systems, in which the
projectile breaks up into two constituents, such as $^6$Li into $d$
and $\alpha$. Thus, the CDCC method is expected to be indispensable to
analyze $^6$Li breakup reactions.

For the CDCC analyses of $^6$Li elastic and inelastic
%%% ver. 8
scattering on various targets, Sakuragi {\it et al.}~\cite{Sakuragi}
%%%
have suggested to reproduce the absolute values of a lot of experimental
data of cross sections very well. They have also resolved the anomalous 
renormalization problem~\cite{Satchler} in folding model potentials of
$^6$Li by showing large dynamical polarization potentials due to breakup
processes. Furthermore the CDCC method has been also applied to the
sequential Coulomb/nuclear breakup via the resonance state of $^6$Li, in
which both Coulomb and nuclear breakup processes were taken into account
consistently~\cite{Hirabayashi}. The Coulomb breakup is of a great
interest as the time-reversed reaction of the radiative capture
reaction,  $d$ + $\alpha$ $\to$ $^6$Li + $\gamma$,
which is one of a key reaction in the nucleosynthesis in the early
Universe or during stellar evolution~\cite{Fowler}. So, it is important
that the method could describe not only the nuclear breakup process but
also the Coulomb breakup one on an equal theoretical footing. 
In spite of their efforts, the CDCC method has not been 
confirmed yet its applicability to breakup to continuum states of
$^6$Li, especially breakup spectra of $^6$Li.
This is a basic interest of the
theoretical framework to describe reaction mechanisms relevant to
breakup processes.  

The $n$ + $^6$Li reactions are important not only from
the basic research interest as shown above but also from the application
point of view. Lithium isotopes will be used as a tritium-breeding
material in $d$-$t$ fusion reactors, and accurate nuclear data are
required for $n$- and $p$-induced reactions. Indeed, IAEA is organizing
a research coordination meeting to prepare nuclear data libraries for
advanced fusion devices, FENDL-3, and the maximum incident energy is set
at 150 MeV to comply fully with the requirements for the IFMIF 
design~\cite{FENDL}. Lithium isotopes are ones of important materials in
these libraries. As a matter of fact, the neutron interaction with
lithium and associated breakup reactions 
are indispensable 
in determining the neutron energy spectra in blankets of fusion
reactors. Thus the tritium breeding ratio, nuclear heating distributions
and radiation damage of structural materials are affected by the 
$n$ + Li reactions significantly.  

%%% ver. 8
{
In spite of the importance of the $n$ + $^6$Li reaction as described
above, experimental data leading to $^6$Li continuum breakup processes
are extremely rare for the neutron energy region above 20
MeV~\cite{Chiba2,Hogue,Hansen}. Furthermore, the statistical model, used
often in evaluation of nuclear data for medium to heavy nuclei, cannot
be applied to the $^6$Li($n$,~$n'$) reactions. As mentioned
in Refs.~\cite{Chiba1,Fukahori}, the mechanisms leading to three- or
four-body final states, $n$ + $d$ + $\alpha$ and $n$ + $n$ + $p$ +
$\alpha$, are important.   
Therefore  more reliable theoretical calculations for the cross sections are
highly desirable. 
}
%
%However it is difficult to calculate cross section data for the
%$^6$Li($n$, $n'$) reaction because the reaction mechanisms
%leading to three- or four-body final states are not understood well
%enough for nuclear data evaluation.
%%%

%%% ver. 8
{ 
In this paper, cross sections for $^6$Li($n$, $n'$)$^6$Li$^*$ $\to$ $d$
+ $\alpha$ reactions are evaluated by using the
microscopic coupled-channels method~\cite{Sakuragi,Takashina-Enyo}, 
in which we adopt microscopic wave functions of $^6$Li with the $d$ +
$\alpha$ model and the complex Jeukenne-Lejeune-Mahaux effective
nucleon-nucleon (JLM) interaction~\cite{JLM} between $n$ and $^6$Li.%
}
%%%
The calculated elastic, inelastic cross sections and continuum neutron
spectra to $^6$Li breakup states show good agreement with the
%experimental data at low neutron incident energies.
{
experimental data at low neutron incident energies, 
where breakup
effects of $^6$Li into $d$ and $\alpha$ are dominant.}
Furthermore we discuss applicability of the approach used here to the
present system for the neutron energy region up to 150 MeV.  

This paper is organized as follows. In Sec.~\ref{sec:Formulation}, we
describe the CDCC method including calculations of $^6$Li wave functions
and coupling potentials with the JLM interaction. In
Sec.~\ref{sec:results} and Sec.~\ref{sec:discussion}, we present the
calculated results and discuss the applicability, respectively. Finally,
we give a summary in Sec.~\ref{sec:summary}.

%%%%%%%%%%%%%%%%%%%%%%%%%%%%%%%%%%%%%%%%%%%%%%%%%%%%%%%%%%%%%%%%%%%
%                            Theory                               %
%%%%%%%%%%%%%%%%%%%%%%%%%%%%%%%%%%%%%%%%%%%%%%%%%%%%%%%%%%%%%%%%%%%
\section{Formulation}
\label{sec:Formulation}

The $n$ + $^6$Li scattering system is described as a $n$ + $d$ +
$\alpha$ three-body model, and the Schr\"{o}dinger equation 
is written by
\begin{eqnarray}
\left[K_{R} + \sum_{j\in^6{\rm Li}}v_{j0}
 + H_{^6{\rm Li}}-E\right]\Psi(\mbold{\xi},\mbold{R})&=&0, \label{hamil-3}
\label{Sch-three}
\end{eqnarray}
where $\mbold{R}$ and $\mbold{\xi}$ represent a coordinate between $n$
and the center-of-mass of $^6$Li and a set of internal coordinates of
$^6$Li, respectively. $K_{R}$ is a kinetic energy operator associated
with $\mbold{R}$, $H_{^6{\rm Li}}$ is the internal Hamiltonian of
$^6$Li, and an effective interaction between $j$th nucleon in $^6$Li and
$n$ is represented by $v_{j0}$.

The total wave function with the total angular momentum $J$
and its projection $M$ on $z$-axis, $\Psi_{JM}$, is expanded
in terms of the orthonormal set of eigenstates of $H_{^6\rm Li}$.
Here, microscopic wave functions of $^6$Li based on the
$d$-$\alpha$ cluster model are written as 
\begin{eqnarray}
 \Phi_{\ell}^{Im}(\mbold{\xi})
  &=&
    \psi_{\ell}^{Im}(\mbold{r})
   \varphi(\alpha)
   \varphi(d),
\end{eqnarray}
where $\mbold{r}$ is the relative coordinate between $d$ and
$\alpha$. The $d$-$\alpha$ relative wave function with the angular
momentum $\ell$, total spin $I$ and its projection $m$ 
on $z$-axis, $\psi_{\ell}^{Im}$, can be described as 
\begin{eqnarray}
 \psi_{\ell}^{Im}(\mbold{r})
  &=&
  \phi_{\ell}^I(r)
  \left[i^\ell
   Y_{\ell}(\Omega_r)\otimes\eta_d
  \right]_{Im},
\end{eqnarray} 
with the spin of deuteron, $\eta_d=1$. The internal ground state
wave functions of $d$ and $\alpha$, $\varphi(d)$ and
$\varphi(\alpha)$, are assumed as inert core. 
In the present calculation, we use the same internal model Hamiltonian
of the $d$-$\alpha$ system, $H_{^6\rm Li}$, adopted in
Ref.~\cite{Sakuragi}, which are fixed to reproduce the experimental data
of the ground state (-1.47 MeV) and $3^+$-resonance state (0.71 MeV) as
well as low-energy part of $d$-$\alpha$ scattering phase shifts.
{ 
The ground state is assumed to be a pure $S$ state ($\ell=0$), and 
}
the calculated wave functions well reproduce not only the elastic
electron scattering form factor but also the inelastic one to the
$3^+$-resonance state~\cite{Sakuragi}. 
In this framework, we take into account the totally
antisymetrized effect by assuming the orthogonal condition
model~\cite{OCM}.  

In the CDCC method, the relative wave functions of $^6$Li for the ground
and continuum states are described as a finite number of discretized
states denoted by 
\begin{eqnarray}
\hat{\psi}_{i\ell}^{Im}(\mbold{r})=\hat{\phi}_{i\ell}^I(r)
\left[i^{\ell }Y_{\ell}(\Omega_r)\otimes\eta_d\right]_{Im}
\,(i=0\mbox{--}N),
\end{eqnarray}
whose energies $\epsilon_{i\ell}^I$ are given by
\begin{equation}
\epsilon_{i\ell}^I
   =  \langle  \hat{\psi}_{i\ell}^{Im} (\mbold{r})
     | H_{^6\rm Li} |
     \hat {\psi}_{i\ell}^{Im} (\mbold{r})
     \rangle_{\mbold{r}}.
\end{equation}
As the discretization approach, we adopt the pseudo-state
method~\cite{Moro1,Matsumoto,Egami} here.  
The advantage of the pseudostate method is that if there are resonances
in its excitation spectrum, we can describe discretized continuum states
with a reasonable number of the basis functions, without distinguishing
the resonance states from non-resonant continuous states as mentioned in
Ref.~\cite{Matsumoto}. As well known, $^6$Li has three resonances in
$\ell=2$. Therefore the pseudostate method is very
useful for analyses of $^6$Li breakup reactions.

In the pseudostate method, we diagonalize $H_{^6{\rm Li}}$  in a space
spanned by a finite number of $L^2$-type basis functions for $r$,
$\{\varphi_{j\ell}\}$, and discretized wave functions for the
radial part, $\{\hat{\phi}_{i\ell}^{I}\}$ are obtained by
\begin{eqnarray}
 \hat{\phi}_{i\ell}^{I}(r)&=&\sum_{j=1}^{j_{\rm max}}
  A_{j\ell}^{I}\varphi_{j\ell}(r).
\end{eqnarray}  
As the basis functions, the complex-range Gaussian basis
functions~\cite{Matsumoto,H-Ka-Ki} are adopted, and we include $\ell=0$
and 2 for the relative angular momentum of $d$-$\alpha$: 
$^1S$ represents the $S$-wave state with $I=1$, 
and $^1D$, $^2D$, and $^3D$ correspond to the $D$-wave state with $I=1$,
2, and 3, respectively.
The number of states $N$ is decided as involving all open
channels for each incident energy.   

After the discretization and the truncation of $d$-$\alpha$ continuum,
$\Psi_{JM}$ is reduced to an approximate one,
\begin{eqnarray}
 &&\Psi^{\rm CDCC}_{JM} =
  \sum_{L}
 {\cal Y}_{JM}^{01L}
 \hat{\phi}_{00}^{1}(r)
 \hat{\chi}_{\gamma_0}(\hat{P}_{\gamma_0},R)/R
 \nonumber \\
&& +
\sum_{i=1}^{N}
\sum_{\ell}
\sum_{I}
\sum_{L}  
{\cal Y}_{JM}^{\ell IL}\hat {\phi}_{i\ell}^I (r)
\hat{\chi}_{\gamma}(\hat{P}_\gamma,R)/R,
\label{appro-expansion}
\end{eqnarray}
with
\begin{eqnarray}
 {\cal Y}_{JM}^{\ell IL}=
\left[[i^{\ell }Y_{\ell}(\Omega_r)\otimes\eta_d\right]_I
\otimes i^L Y_{L}(\Omega_R)]_{JM}\varphi(\alpha)\varphi(d),
\end{eqnarray}
where $\gamma_0=(0,0,1,L,J)$ and $\gamma=(i,\ell,I,L,J)$ represent the
elastic channel and breakup channels, respectively. The
expansion-coefficient $\hat{\chi}_{\gamma}$ in
Eq.~(\ref{appro-expansion}) represents the relative motion between
$n$ and $^6$Li, and $L$ is the orbital angular momentum regarding
$\mbold{R}$. The relative momentum $P_{\gamma}$ is determined by the 
conservation of the total energy,
\begin{eqnarray}
E&=&\hat{P}_{\gamma}^{2}/2\mu+\epsilon_{i\ell}^I,
\end{eqnarray} 
with $\mu$ the reduced mass
of the $n$ + $^6$Li system. Multiplying Eq.~(\ref{Sch-three})
by $[\hat{\phi}_{i\ell}^I(r){\cal Y}_{JM}^{\ell I L}]^*$ from
left, one can obtain a set of coupled
differential equations for $\hat{\chi}_{\gamma}$, called the CDCC
equation. Solving the CDCC equation under the appropriate asymptotic
boundary condition, we can obtain the elastic and discrete breakup
$S$-matrix elements. Details of the formalism of the CDCC method are
shown in Ref.~\cite{Kamimura}. 

\begin{figure*}[htbp]
 \includegraphics[width=0.35\textwidth,clip]{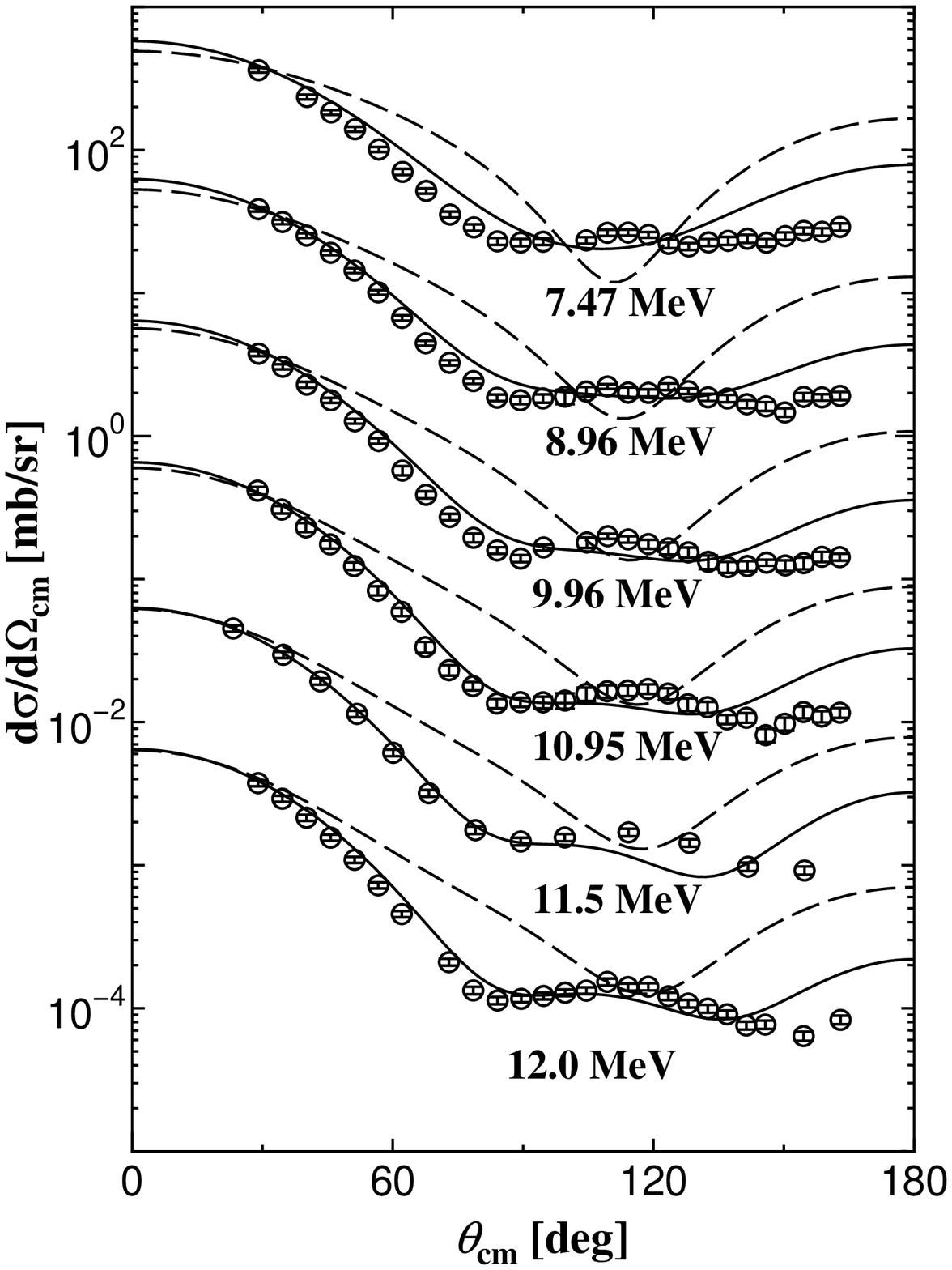}
 \includegraphics[width=0.35\textwidth,clip]{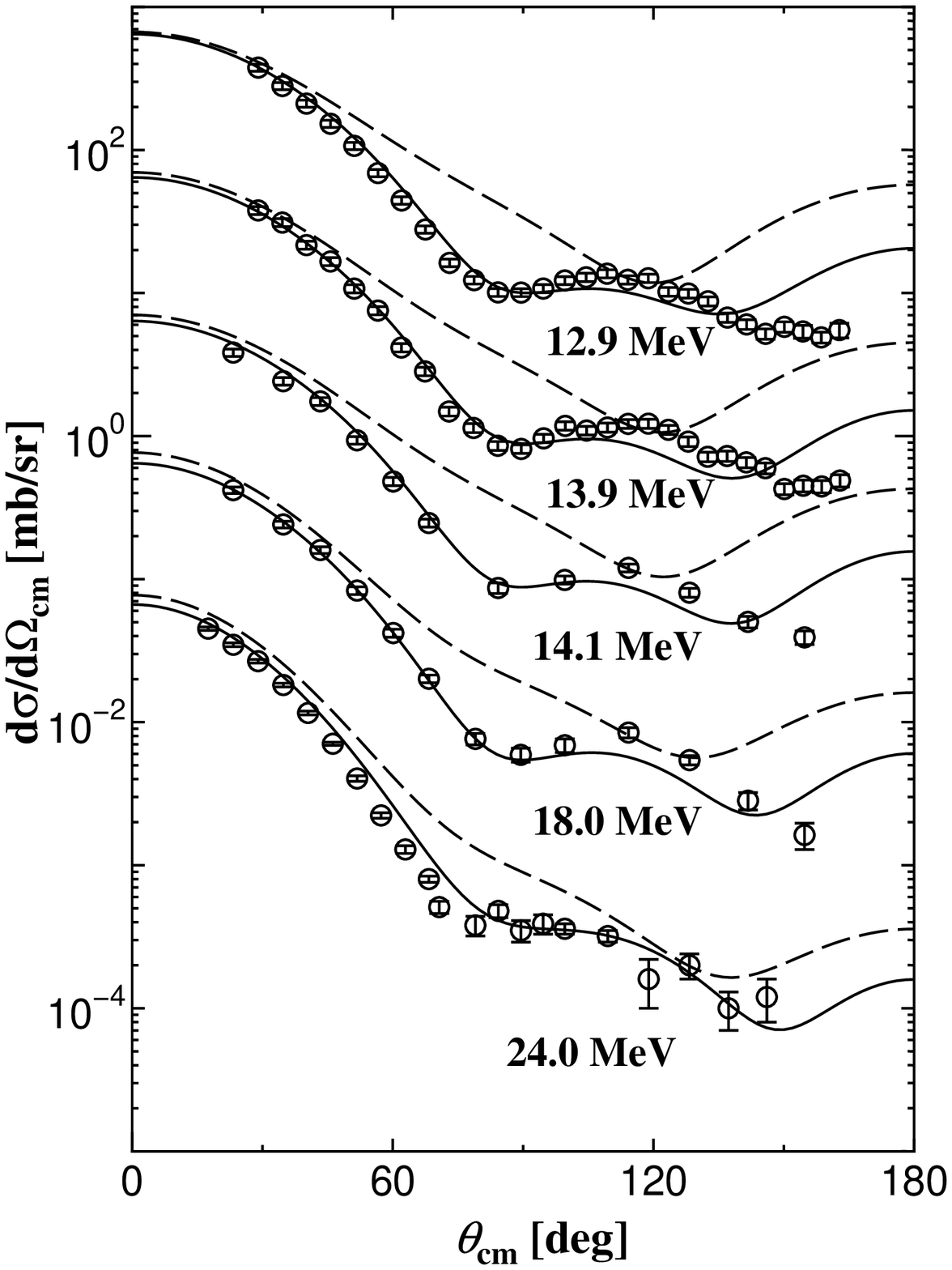}
 \caption{Angular distribution of the elastic differential cross section
 of $n$ + $^6$Li scattering for incident energies between 7.47 and 24.0
 MeV. The solid and dashed lines correspond to the result with and without
 couplings to breakup states of $^6$Li, respectively.
 Experimental data are taken from Refs.~\cite{Chiba1,Hogue,Hansen}.
 The data are subsequently shifted downward by a factor of
 $10^{-1}$-$10^{-5}$ from 8.96 MeV to 12.0 MeV on the left panel, and
 $10^{-1}$-$10^{-4}$ from 13.9 MeV to 24.0 MeV on the right panel,
 respectively.}    
 \label{elastic}
\end{figure*}

For the calculation of diagonal and coupling potentials in
the CDCC equation, we use the complex Jeukennd-Lejeune-Mahaux effective
nucleon-nucleon (JLM) interaction~\cite{JLM} based on a single folding
model. The JLM interaction can be easily applied to coupled-channels
calculations of nucleon-nucleus systems, such as
Refs.~\cite{Takashina-Enyo,Keeley-Lapoux}.   
This interaction has energy and density dependence, and forms a
Gaussian with both real and imaginary parts;
\begin{eqnarray}
 v_{j0}(R_{j0};\rho,E)
  &=&
  \lambda_{v}V(\rho,E)\exp\left(-R_{j0}/t_R^2\right)\non\\
 &&{}+i\lambda_wW(\rho,E)\exp\left(-R_{j0}/t_I^2\right),
  \label{Eq:JLM}
\end{eqnarray}  
where $R_{j0}$ is a coordinate of a nucleon in $^6$Li and $n$. 
%%% ver. 8
{ 
The parameters, $t_R$, $t_I$, and $\lambda_v$,
are taken as the same ones used in the original paper~\cite{JLM},
$t_R=t_I=1.2$ and $\lambda_v=1.0$. Meanwhile the normalization
for the imaginary part, $\lambda_w$, is optimized to reproduce the
elastic cross sections, because the strength corresponding to the loss
of flux depends on the model space considered in the calculation.}
%%%
Details for strengths of $V(\rho,E)$ and $W(\rho,E)$ are shown in
Ref.~\cite{JLM}. 

Using the JLM interaction, the diagonal and coupling potentials,
$V_{\gamma\gamma'}$, are obtained by 
\begin{eqnarray}
  V_{\gamma \gamma'}(R)   
 &=&\int \rho_{\gamma\gamma'}(\mbold{s},\Omega_R)
  v_{j0}(E,\bar{\rho},\mbold{R}_{j0})d\mbold{s}d\Omega_R, 
\label{coupling}
\end{eqnarray}
where transition densities, $\rho_{\gamma\gamma'}$,
and averaged matter density, $\bar{\rho}$, of $^6$Li between
$\gamma$ and $\gamma'$ are defined by
\begin{eqnarray}
 \rho_{\gamma\gamma'}(\mbold{s},\Omega_R)
  =\langle {\cal Y}_{JM}^{\ell IL}\hat {\phi}_{i\ell}^{I}
     | \sum_{j\in^6{\rm Li}}\delta(\mbold{s}-\mbold{s}_j)| 
     {\cal Y}_{JM}^{\ell'I'L'}
     \hat{\phi}_{i'\ell'}^{I'}
     \rangle_{\mbold{\xi}},\nonumber\\
\end{eqnarray}
and
\begin{eqnarray}
 \bar{\rho}(s)=
  \frac{1}{2}\int \left\{\rho_{\gamma\gamma}(\mbold{s},\Omega_R)
	+\rho_{\gamma'\gamma'}(\mbold{s},\Omega_R)
	\right\}d\Omega_sd\Omega_R,
\end{eqnarray}
respectively. Here $\mbold{s}_j$ is a coordinate of $j$th particle in
$^6$Li relative to the center-of-mass of $^6$Li.

\begin{figure}[htbp]
 \includegraphics[width=0.4\textwidth,clip]{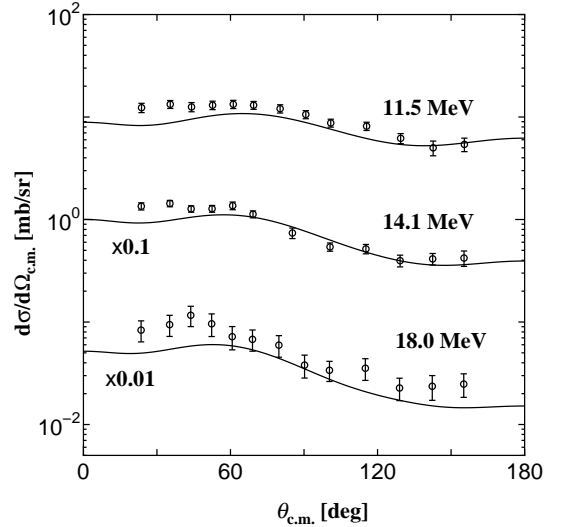}
 \caption{Neutron inelastic scattering angular distribution to the
 $3^+$-resonance state 0.71 MeV above the $d$-$\alpha$ threshold. The
 experimental data are  taken from 
 Ref.~\cite{Chiba1}. The data for 14.1 (18.0) MeV are shifted by a factor of
 0.1 (0.01).}    
 \label{inelastic}
\end{figure}

\begin{figure*}[htbp]
 \includegraphics[width=0.8\textwidth,clip]{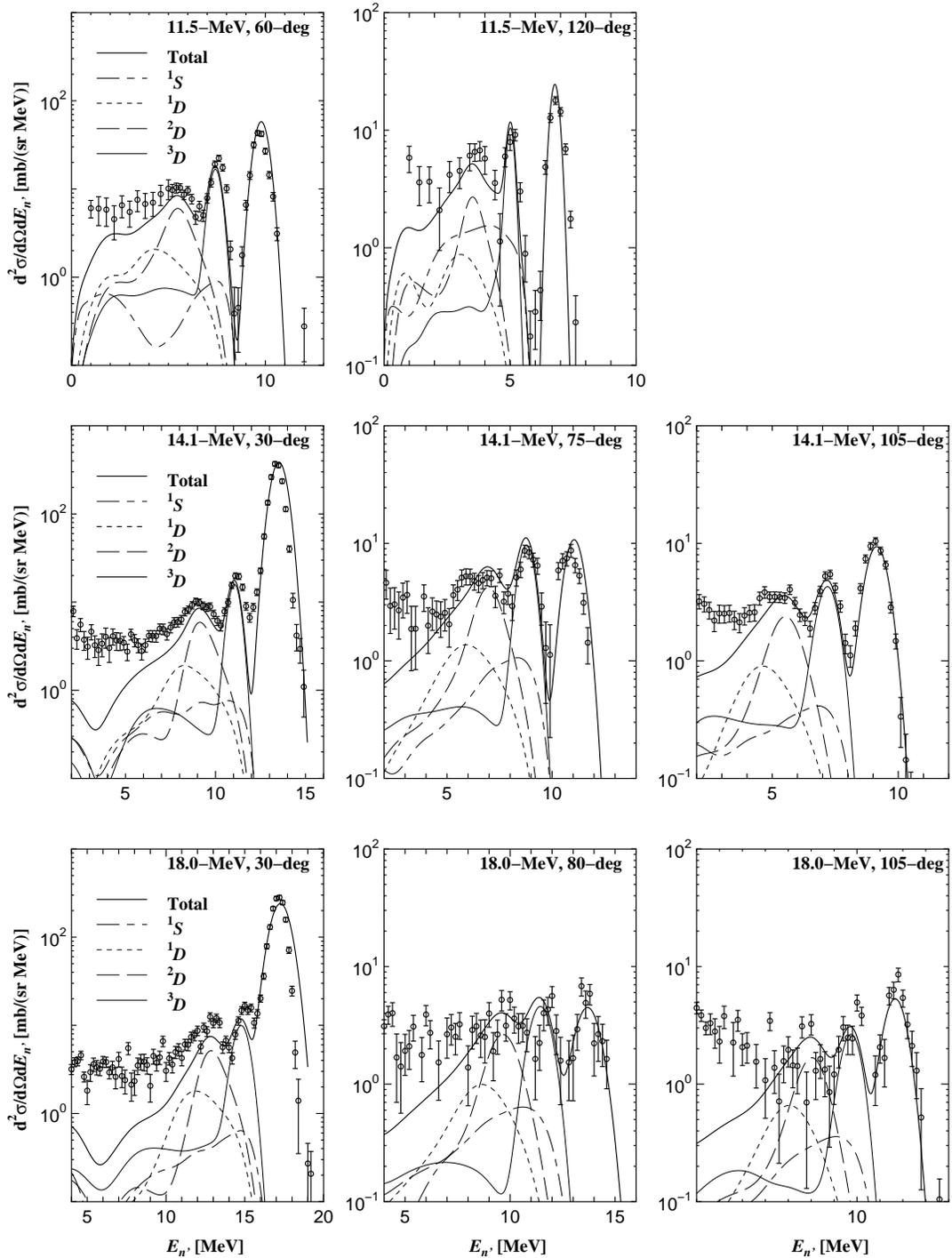}
 \caption{The neutron spectra calculated by the CDCC
 method with the JLM interaction comparing with measured data at
 selected angular points in the laboratory system. The experimental data
 are taken from Ref.~\cite{Chiba1}.} 
 \label{neutron-spectrum}
\end{figure*}

\section{Results}
\label{sec:results}

%%% ver. 8
Figure \ref{elastic} shows the differential elastic cross sections of $n$
+ $^6$Li for incident energies between 7.47 and 24.0 MeV. One sees that
the results of the CDCC calculation represented by the solid lines are
in good agreement with the experimental data. The dashed lines represent
results of a single channel calculation, in which couplings to breakup
states are omitted. It is found that breakup effects shown by the
difference between the dashed and solid lines are significant to
reproduce the angular distributions of the elastic scattering.
%%% ver. 8
{
For all incident energies, we take $\lambda_w=0.1$ to reproduce
the data. The main component of the
flux loss from the elastic channel is due to breakups of $^6$Li to $d$
and $\alpha$, which can be taken into account in the CDCC calculation
directly. The other effects of the flux loss, such as the excitation of
$\alpha$ and $d$ breakup, are not so large because of low
incident energies. Therefore the strength of $\lambda_w$ becomes very
small. Here, it should be noted that the single channel calculation
cannot reproduce the experimental data if any values of $\lambda_w$
are taken.
} 
%%%

Figure~\ref{inelastic} shows the angular distributions to the first
excited $3^+$ state of $^6$Li for $E_n=$11.5, 14.1, and 18.0 MeV. The 
theoretical cross sections are calculated by integrating the breakup
cross section to $3^+$-continuum for the resonance energy region. One
sees that the CDCC calculation can also reproduce the inelastic cross
section.

In Fig.~\ref{neutron-spectrum}, the calculated neutron spectra are
compared with the experimental data at selected angles in the laboratory
frame and incident energies. 
%%%
Components of $^1S$, $^1D$, $^2D$, and $^3D$ are represented by the
dash-dotted, dashed, dotted and thin solid lines, respectively, and
these results are broadened by considering the finite resolution of the
experimental apparatus~\cite{Chiba1}. 
%%%
Three peaks in the experimental data represent the elastic, inelastic to
the $3^+$-resonance, and $2^+$-resonance components, respectively, from
higher neutron energies. 
%%% 
The CDCC calculation gives a good agreement with experimental data at
the higher neutron energy region. On the other hand in the low neutron
energy region, which corresponds to high exited states of $^6$Li,
the calculated cross section underestimates the experimental data for
all energies and angles.
%The result is considering due to the fact that experimental
The result indicates that experimental
data contain contribution from the ($n$, 2$n$) reaction corresponding to
a four-body breakup reaction $^6$Li($n$, $nnp$)$\alpha$, as mentioned in
Ref.~\cite{Chiba1}. 
%%%
In the present calculation, the four-body breakup
effect is not taken into account directly, and the effect is treated as
a absorption effect on the elastic cross section.

\begin{figure}[htbp]
 \includegraphics[width=0.37\textwidth,clip]{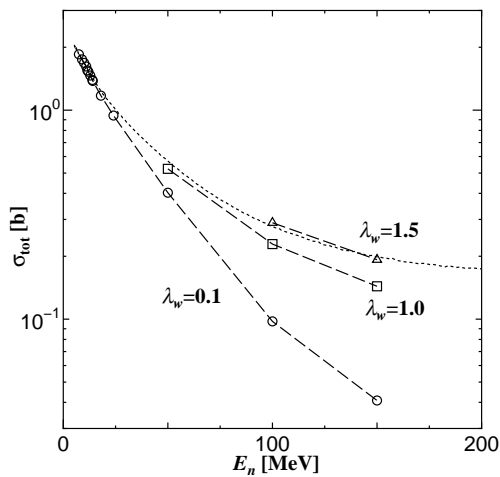}
 \caption{Neutron energy dependence of the total cross section of $n$ +
 $^6$Li scattering. The dotted line represents the experimental
 data~\cite{n-tot-exp}. The open circles, squares, and triangles
 corresponds to results with $\lambda_w=0.1$, 1.0, and 1.5, respectively.} 
 \label{total}
\end{figure}

\section{Discussion}
\label{sec:discussion}

Theoretically the CDCC calculation with the JLM interaction can estimate
cross sections of the $n$ + $^6$Li scattering for neutron energies up to 160
MeV, which is the maximum applicable energy suggested in the original
paper~\cite{JLM}. However  for higher incident energies ($>30$ MeV)
there is no experimental data of the elastic cross section, which
are needed to optimize the normalization factor of the imaginary part. 
One of the optimization methods is to fit the neutron total
cross section data of $^6$Li, which have been measured 
up to 560 MeV~\cite{n-tot-exp}. Figure~\ref{total} shows 
the measured total cross section comparing with the calculated ones with
$\lambda_w=$0.1, 1.0, and 1.5 represented by open circles,
squares, and triangles, respectively. One sees that large values
for $\lambda_w$ are required with respect to increasing the neutron
energy. Although the energy dependence for $\lambda_w$ is
rationalized because absorption effects become strong as increasing the
energy, the optimized value at 150 MeV is much large about
$\lambda_w$=1.5. This normalization problem would be resolved by
analyzing $p$ + $^6$Li elastic cross sections, which have been
measured at higher incident energies, within the present framework.
Furthermore for higher incident neutron energies four-body breakup
effects of $^6$Li($n$,~$nnp$)$\alpha$ are not negligible.  
Therefore we require the four-body CDCC calculation,
which has been successful for analyses of $^6$He breakup
reactions~\cite{Matsumoto3,Matsumoto4,THO-CDCC,4body-CDCC-bin,Matsumoto5},
in the higher energy region.

\section{Summary}
\label{sec:summary}

We analyze $^6$Li($n$, $n'$)$^6$Li$^*$ $\to$ $d$ + $\alpha$ reactions by using
the continuum-discretized coupled-channels method with the complex 
Jeukenne-Lejeune-Mahaux effective nucleon-nucleon interaction. 
In the present analysis, it is found that the elastic cross
sections for incident energies between 7.47 and 24.0 MeV can be
reproduced by the present analysis with one normalization parameter 
for the imaginary part of the JLM interaction ($\lambda_w=0.1$), and
breakup effects on the elastic cross section are significant.
Furthermore the calculated inelastic cross section to the
$3^+$-resonance state and neutron spectra are also good agreement
with the experimental data systematically.
%%% ver. 8
{
Thus, the CDCC method with the JLM interaction is expected to be
indispensable for the data evaluation of the $^6$Li($n$, $n'$)
reactions, and the advantage %of the CDCC method 
is to obtain not only elastic and inelastic cross sections but also
neutron spectra within the same framework.  
} 
%%%

For higher neutron incident energies, we have also discussed the
applicability of the present microscopic approach. It is found that
the required normalization factor $\lambda_w$ is much large,
$\lambda_w=1.5$ for 150 MeV, from the analyses of the total cross
section. In order to investigate the normalization problem we should
analyze $p$ + $^6$Li elastic scattering, which have been measured at
higher incident energies. Furthermore the four-body CDCC calculation is
required to describe the four-body breakup process of $^6$Li into $n$ +
$p$ + $\alpha$, which becomes significant for the higher energy. These
results will be reported in a forthcoming paper.  

\section*{Acknowledgments}

We would like to thank the members of Nuclear Theory Group at Hokkaido
University. This work was supported by JSPS AA Science Platform Program.

%---------------------------------------------------------------------
%%%%%%%%%%%%%%%%%%%%%%%%%%%%%%%%%%%%%%%%%%%%%%%%%%%%%%%%%%%%%%%%%%%
%                          References                             %
%%%%%%%%%%%%%%%%%%%%%%%%%%%%%%%%%%%%%%%%%%%%%%%%%%%%%%%%%%%%%%%%%%%

\end{document}